# The Switch, the Ladder, and the Matrix: Models for Classifying AI Systems

Jakob Mökander,[1] Margi Sheth,[2] David S. Watson,[3] Luciano Floridi[1,4]

[1] Oxford Internet Institute, University of Oxford, 1 St Giles', Oxford, OX1 3JS
[2] R&D Data Office, AstraZeneca PLC, 1 Francis Crick Avenue, Cambridge CB2 0AA, UK
[3] Department of Statistical Science, University College London, London WC1E 6BT, UK
[4] Department of Legal Studies, University of Bologna, Via Zamboni 33, Bologna, 40126, Italy

Email for correspondence: jakob.mokander@oii.ox.ac.uk

## Abstract

Organisations that design and deploy artificial intelligence (AI) systems increasingly commit themselves to high-level, ethical principles. However, there still exists a gap between principles and practices in AI ethics. One major obstacle organisations face when attempting to operationalise AI Ethics is the lack of a well-defined material scope. Put differently, the question to which systems and processes AI ethics principles ought to apply remains unanswered. Of course, there exists no universally accepted definition of AI, and different systems pose different ethical challenges. Nevertheless, pragmatic problem-solving demands that things should be sorted so that their grouping will promote successful actions for some specific end. In this article, we review and compare previous attempts to classify AI systems for the purpose of implementing AI governance in practice. We find that attempts to classify AI systems found in previous literature use one of three mental models: *the Switch*, i.e., a binary approach according to which systems either are or are not considered AI systems depending on their characteristics; *the Ladder*, i.e., a risk-based approach that classifies systems according to the ethical risks they pose; and *the Matrix*, i.e., a multi-dimensional classification of systems that take various aspects into account, such as context, data input, and decision-model. Each of these models for classifying AI systems comes with its own set of strengths and weaknesses. By conceptualising different ways of classifying AI systems into simple mental models, we hope to provide organisations that design, deploy, or regulate AI systems with the conceptual tools needed to operationalise AI governance in practice.

**Note**:
This is a pre-print. A revised version has accepted for publication in Minds and Machines and can be cited as follows:

## 1    Introduction

Artificial intelligence (AI) is increasingly reshaping societies and transforming economies (AlgorithmWatch, 2019). This is understandable: the delegation of tasks to AI systems holds great promise to improve efficiency, reduce costs, and enable new solutions to complex problems (Taddeo & Floridi, 2018). For example, AI systems can improve health outcomes (Grote & Berens, 2020; Schneider, 2019) and help mitigate environmental risks (Rolnick et al., 2019; Vinuesa et al., 2019). However, the use of AI systems is coupled with ethical challenges. A particular AI system may be poorly designed, leaving individuals and groups vulnerable to poor quality outcomes, bias and discrimination, and invasion of privacy (Leslie, 2019). Further, AI systems can enable human wrongdoing, reduce human control, and erode human self-determination (Tsamados et al., 2020). At the same time, fear and misplaced concerns could hamper the adoption of well-designed AI systems, thereby leading to significant social opportunity costs (Cookson, 2018). These and other similar ethical challenges cannot be ignored if one wishes to reap the benefits brought by AI systems.

Many governments, research institutes, and NGOs have proposed ethical principles that provide normative guidance to organisations that design and deploy AI systems (Fjeld, 2020; Jobin et al., 2019).[1] Although differing in terminology, these guidelines tend to converge on five principles: beneficence, non-maleficence, autonomy, justice, and explicability (Floridi & Cowls, 2019). In parallel, numerous organisations have adopted AI ethical principles of their own (de Laat, 2021). Notable examples include Google (2018), Microsoft (2019), IBM (Cutler et al., 2018), BMW Group (2020) and AstraZeneca (2020). Collectively, these efforts constitute a step in the right direction. However, the adoption of (and subsequent adherence to) AI ethical principles remains voluntary (Cath et al., 2018) and often unchecked. Moreover, the industry lacks both incentives and useful tools to translate abstract principles into verifiable criteria (Morley et al., 2020; Raji et al., 2020).

Legislation has only recently begun to change this picture. The *Artificial Intelligence Act* (published by the European Commission on 24 April 2021) was the first comprehensive legislative framework for AI proposed by any major global economy. During the last year, many other countries and regions have followed suit. For example, the U.S. Senate and House are currently considering the *Algorithmic Accountability act of 2022* (Office of U.S. Senator Ron Wyden, 2022). Yet whether and when this bill will pass into law remains uncertain.

---

[1] Recent and influential contributions include the European Commission's Ethics Guidelines for Trustworthy AI (AI HLEG, 2019), the IEEE's principles for Ethically Aligned Design (IEEE, 2019), and the OECD's Recommendation of the Council on Artificial Intelligence (OECD, 2019).

Outside the domain of hard governance,[2] however, much has already been done to bridge the gap between principles and practice in AI ethics (Ibáñez & Olmeda, 2021; Morley et al., 2021; Schiff et al., 2021b). Institutions have produced detailed assessment lists (AI HLEG, 2020; Reisman et al., 2018); researchers have developed translational tools like model cards (Mitchell et al., 2019) and datasheets (Gebru et al., 2018; Holland et al., 2018); industry-specific initiatives have drafted standardised protocols and reporting guidelines for the use of AI systems (Cruz Rivera et al., 2020; Liu et al., 2020), and private organisations are increasingly adopting ethics-based auditing procedures (Brundage et al., 2020; Deloitte, 2020; Mökander et al., 2021a; PwC, 2019; Sandvig et al., 2014). All these efforts serve the overarching purpose of enabling effective AI governance, i.e., to provide organisations with the *tools* needed to ensure that the AI systems they design are legal, ethical, and technically robust.

However, the lack of a clear *material scope[3]*—that is, to which technological systems the ethical and legal considerations may or may not apply—continues to make it difficult to implement and enforce AI governance frameworks in practice (Kritikos, 2019; Scherer, 2016). For example, in a recent industry case study (which was based on interviews with managers and software developers), we found that one of the main challenges organisations face when attempting to operationalise AI governance is the lack of procedures for demarcating the material scope of such initiatives (Mökander & Floridi, 2022). As a result, organisations are often unable to produce an inventory of the AI systems they develop or use. Despite such difficulties, however, both organisations that commit themselves to AI ethics principles and regulators that develop AI governance frameworks inevitably face the question 'to which systems and processes ought these additional layers of governance to apply?' (Aiken, 2021).

Of course, there is no one way to demarcate the material scope of AI governance. Different AI systems pose different ethical and legal challenges (Oxborough et al., 2018). Moreover, AI systems are often embedded in larger socio-technical systems (Lauer, 2020; van de Poel, 2020) in which human- and machine-centric processes overlap and co-evolve (Di Maio, 2014; Tam et al., 2017). Admittedly, this ontological underdetermination is not unique to the problem of AI governance. Grouping things into neat categories seldom works, given the messy and continuous boundaries of the natural world (Smith, 2019). However, for the purpose of pragmatic inquiry and practical problem solving, things have to be sorted so that their grouping can promote successful actions for some specific end (Dewey, 1957). In short, every policy needs to define its material scope (Schuett, 2021).[4]

---

[2] Hard governance refers to legally binding obligations (Erdelyi & Goldsmith, 2018). In contrast, soft governance refers to mechanisms that exhibit some flexibility, like cultural norms or financial incentives (Floridi, 2018).
[3] The term "material scope" is standard in the context of the General Data Protection Regulation, see Art. 2, https://gdpr-info.eu/art-2-gdpr/.
[4] We use the term 'material scope' rather than just 'scope' because a given policy can also have other types of scope (such as 'territorial scope').

In this article, we analyse the material scope of existing AI governance frameworks. Through a systematised literature review (Grant & Booth, 2009), we identify and compare previous attempts to classify AI systems for the practical purpose of operationalising corporate AI governance.[5] We find that they follow one of three approaches. According to the *binary approach*, systems either are or are not considered AI systems, depending on their intrinsic characteristics. According to the *risk-based approach*, systems are classified into different categories depending on the types of ethical risks they pose. Finally, according to the *multi-dimensional approach*, various aspects—such as context, data input, and decision-model type—need to be considered when classifying systems. Using mental models (Johnson-Laird, 1983), we call these approaches *the Switch*, *the Ladder*, and *the Matrix*, respectively. In the following sections, we discuss each of these models in detail and provide several concrete examples.

Before proceeding, three limitations help demarcate the scope of this article. First, we do not undertake any normative evaluation of different AI ethics guidelines. Previous research has pointed out that the apparent consensus around high-level principles may hide tensions, for example, in terms of priorities or how concepts like justice or fairness should be interpreted (Schiff et al., 2021a). As highlighted by (Whittlestone et al., 2019), different ethical principles sometimes give rise to tensions for which there are no fixed solutions. Therefore, organisations are expected to strike justifiable trade-offs within the limits of legal permissibility and operational viability. However, we assume *normative clarity*, i.e., that an organisation seeking to implement an AI governance framework has already committed itself to a coherent set of principles, e.g., to the EU or OECD ethical guidelines.

Second, for the purpose of our argument, we do not need to engage with questions concerning *good intent*. Ideally, organisations that commit to AI ethics principles also seek to live by these values. In practice, however, commitments to ethical principles can be undermined by unethical practices like 'ethics blue washing', i.e., making unsubstantiated claims about AI systems to appear more ethical than they are, or 'ethics lobbying', i.e., exploiting ethics to delay or avoid necessary legislation (Ferretti, 2021; Floridi, 2019). While important, these considerations lie outside the scope of this article. Instead, we take as our starting point the premise that good classification of AI systems may facilitate but never guarantee morally good outcomes.

Finally, our review does not encompass abstract definitions of what 'artificial intelligence' really *is*. As is well-known, there exists no universally accepted definition of AI (Wang, 2019).[6]

---

[5] To clarify: In this article, we do not propose any new model for how to classify AI systems. Instead, our focus is on identifying, describing, and evaluating different ways of classifying AI systems found in the recent literature.

[6] Some researchers use the term AI to refer to specific types of agents, i.e., those displaying some levels of autonomy, adaptability, and problem-solving capacity (Legg & Hutter, 2007). Others take AI to demarcate the set of computational techniques that are designed to approximate cognitive tasks (*US National Defence*

Discussions concerning the merits of different universal definitions of AI remain outside this article's scope. Instead, we focus on *classifications* that help organisations implement and enforce their AI governance frameworks. To quote John Dewey (Dewey, 1957), "To have an aim is to limit, select, concentrate, and group". And we have an aim in mind: to unlock the potential of autonomous and self-learning systems to serve as a force for good while managing the ethical challenges they pose.

To describe and discuss different models for classifying AI systems, we rely on the method of levels of abstraction (Floridi, 2008). Abstraction is an appropriate method for analysing and understanding complex phenomena since it allows for the creation of concepts and objects at different levels of thinking and language (van Leeuwen, 2014). Only within a level of abstraction (LoA) can comparison between objects make sense. However, note that this is not a relativist approach: a question is always asked for a purpose, and, for that specific purpose, there is an appropriate LoA that can be compared to others in terms of "fitting" the purpose more or less successfully.

The remainder of this article is structured as follows. In Section 2, we build on previous work to showcase how AI systems can be classified in many different ways based on their technical features, the socio-technical contexts in which they are applied, and on the LoA for which a classification is sought. In Section 3, we argue that, to establish the material scope of AI governance, good classifications of AI systems should be *fit for purpose*, *simple and clear*, and *stable over time*. We then introduce three models for how to classify AI systems. In Sections 4—6, we describe and exemplify the Switch, the Ladder, and the Matrix*, respectively.* In Section 7, we evaluate these models according to the criteria set out in Section 3. Finally, in Section 8, we conclude by discussing how classifying AI systems is an LoA-dependent question. Hence, none of the models discussed in this article should be viewed as applicable absolutely, that is, independently of the choice of the LoA deemed to be most appropriate for the given purpose. Instead, we suggest that the models for classifying AI systems outlined in this article collectively constitute a useful set of tools for technology providers or regulators that wish to clarify the material scope of their AI governance frameworks.

## 2  Conceptualising 'AI systems'

At a high LoA, AI systems can be viewed as interactive, autonomous, and self-learning systems that can perform tasks that would otherwise require human intelligence and intervention to be executed successfully (Floridi et al., 2018). Still, to help AI practitioners understand whether an AI governance framework applies in a particular case, it must be complemented by a classification of AI systems at lower LoAs.

---

*Authorization Act*, 2018). Yet others use the term AI not to describe any technologies at all, but rather the science and engineering of making performant machines (McCarthy, 2007).

Note that the term 'AI system' here indicates that we are talking about a class of systems that differs from others, as opposed to some kind of intelligence that differs from human intelligence (Kostopoulos, 2021). To capture this distinction, a wide range of terms like 'AI-based systems' (Gasser & Almeida, 2017; Saleiro et al., 2018), 'algorithmic systems' (Ananny & Crawford, 2018; Rahwan, 2018), 'automated decision-making systems' (AlgorithmWatch, 2019; Whittaker et al., 2018), and 'autonomous/intelligent systems' (Bryson & Winfield, 2017; IEEE SA, 2020) are often used interchangeably in the existing literature. For the sake of simplicity, we shall use the term 'AI system' consistently throughout this article. In doing so, we follow the (OECD, 2020) and the Alan Turing Institute (Leslie, 2019). However, nothing hinges on the choice.

Previous work has shown that AI systems can be classified according to several different dimensions. A distinction is often made between narrow and general AI systems (Russell et al., 2015). While *narrow* (or weak) AI refers to systems that can outperform humans on specific cognitive tasks, *general* (or strong) AI refers to systems that demonstrate human-level intelligence across a broad range of cognitive tasks (Goldstein, 2018). So defined, most current AI systems are narrow, although there is a growing body of research on transfer learning (Weiss et al., 2016) and meta-learning (Vanschoren, 2018) explicitly devoted to building models that generalize across tasks.

Another distinction is often made between different AI paradigms. While *symbolic* approaches are based on logic programming and symbol manipulation, *adaptive* methods rely on statistical techniques to solve specific problems without being explicitly programmed to do so (Russell & Norvig, 2015). This latter class includes machine learning algorithms, such as decision trees and deep neural networks (Samoili et al., 2020). However, the two approaches are not necessarily mutually exclusive. So-called *hybrid architectures* attempt to combine the large-scale learning abilities of neural networks with symbolic knowledge representation (Marcus, 2020).

Within the realm of adaptive approaches, practitioners distinguish between supervised-, unsupervised-, and reinforcement learning. *Supervised learning* involves inferring a relationship from inputs to outputs, e.g., classifying image labels from pixels or predicting economic demand from time-series data (Hastie et al., 2009). In contrast, *unsupervised learning* is about finding patterns (e.g., clusters or latent variables) hidden in collections of unlabelled data without any predetermined target (Frankish & Ramsey, 2014). Finally, *reinforcement learning* occurs when an agent attempts to maximise rewards by interacting within some structured environment (Sutton & Barto, 2018). Policies are gradually improved through repeated trials, as when AlphaGo (Silver et al., 2016) became the world's greatest master of the ancient Chinese game 'Go' by playing against itself millions of times.

AI systems can also be classified with respect to the type of cognitive tasks they attempt to emulate (Feigenbaum & Feldman, 1963). Traditionally, AI research has focused on the following problem domains: *perception*, i.e., the ability to transform sensory inputs into usable information;

*reasoning*, i.e., the capability to solve problems; *knowledge*, i.e., the ability to represent and understand the world; *planning*, i.e., the capability of setting and achieving goals; and *communication*, i.e., the ability to understand and produce language (Corea, 2019). An alternative and complementary way of classifying AI systems is based on the type of analytics they perform. These include *descriptive analytics* (what happened?), *diagnostic analytics* (why did something happen?), *predictive analytics* (what is going to happen?), *prescriptive analytics* (what should happen?), and *automated analytics* (performing actions) (Corea, 2019).

Here, it is worth mentioning that in the proposed European regulation (the *Artificial Intelligence Act*, or AIA), AI systems are defined by the combination of the technical approaches that underpin a system and the cognitive tasks that the system is designed to perform (European Commission, 2021b). More specifically, in Annex 1 to the AIA, AI systems are defined as:

> software that [i] is developed with one or more of the [following] techniques and approaches: (a) Machine learning approaches, […]; (b) Logic- and knowledge-based approaches, […]; and (c) Statistical approaches, […], and [ii] can, for a given set of human-defined objectives, generate outputs such as content, predictions, recommendations, or decisions influencing the environments they interact with.

The definition of AI system provided in the AIA is broad by any standard.[7] Further, the diverse nature of the technical approaches it encapsulates—and the wide range of applications they enable —shows that it is often necessary to consider LoA-dependent factors when classifying AI systems. In practice, AI systems are not isolated technologies but integrated into larger socio-technical systems that encompass organisations, people, infrastructures, and processes (Chopra & Singh, 2018). Put differently, information processing—from the collection of input data to the final decision or classification—typically consists of several interconnected (and often iterative) steps performed by both human operators and computational systems (Chen & Golan, 2016). Hence, the decisions made by AI systems are never just a reflection of their technical properties but also of the socio-technical environment surrounding their use (Eubanks, 2019).

However, all technical artefacts (AI systems included) are value-laden insofar as they alter the cost-benefit ratio of the actions undertaken by humans and thus influence their decision-making (Danaher, 2012). Moreover, some AI systems can adapt their behaviour based on external inputs and evolve over time. This ability of AI systems 'to learn', i.e., to update continuously their internal decision-making logic, is one of the reasons why it is difficult to assign accountability when harm occurs (Burrell, 2016; Floridi, 2016). From a socio-technical perspective, it is this combination of relative autonomy and learning skills that underpin both beneficial and problematic uses of AI systems. To determine the risk level, or harm potential, it is often necessary to take several factors into account, including *data access*, i.e., the extent to which a system has complete and accurate knowledge about its environment; *model stability*, i.e., the extent to which a system may alter its own control structure to perform its task; and *goal freedom*, i.e., the extent to which the system's goals are known and stable.

---

[7] For a full discussion concerning the limitations of the material scope of the proposed European legislation (as currently defined), please see our article *Conformity Assessments and Post-market Monitoring: A Guide to the Role of Auditing in the Proposed European AI Regulation* (Mökander et al., 2021b).

To summarise, previous work in the field suggests that separating AI systems from other systems is a LoA-dependent, multi-variable problem, and, as we shall see in Sections 4—6, this problem can be approached in different ways. Before discussing these different approaches in greater detail, let us explore the needs of an effective classification system and establish criteria for the same.

## 3 Criteria for good classifications of AI systems

To implement AI governance in practice, the concept 'AI systems' must be operationalised for three main reasons. First, organisations are under constant pressure to innovate. By having a clearly defined material scope for their AI governance, organisations can take care not to unduly burden systems or projects from which no AI-specific risks arise (AIEIG, 2020).

Second, governance is most effective when rules and norms are applied fairly, transparently, and consistently (Hodges, 2015). Without a shared understanding of what constitutes AI systems within a specific organisation, systems, and processes (henceforth *use cases*) are likely to be subject to additional scrutiny only on an *ad hoc* basis. Such a procedure undermines the legitimacy of the AI governance framework in question and hamper its ability to systematically identify ethical risks.

Third, and most importantly, not all ethical risks that organisations face stem from the use of AI systems. The inherent technical opacity of AI systems, for example, is often dwarfed by the opacity stemming from state secrecy or intellectual property rights (Burrell, 2016). Further, human decision-makers also make mistakes and produce discriminatory or inconsistent outcomes (Kahneman, 2011). As a result, organisations already have processes in place to oversee human decision-making and enforce commitments related to environmental sustainability, corporate social responsibility, and data management (e.g., in line with the *General Data Protection Regulation*). Therefore, a well-defined material scope for AI governance frameworks should complement or refine existing governance structures, not duplicate, or generate inconsistencies within them.

What constitutes a good classification of AI systems? Drawing on best practices from the legal tradition (Baldwin & Cave, 1999) and attempts to create *working definitions* within the philosophy of science (Carnap, 1950), we argue that good classifications should be:

1) *Fit for purpose*: the classification should help organisations demarcate the material scope of AI governance in ways that are neither over- nor underinclusive. A classification of AI systems is overinclusive when it includes systems or use cases that do not require additional oversight with respect to the normative goals of the AI governance framework in question (e.g., items already covered under data management and governance principles). In contrast, a classification is underinclusive when systems or use cases that pose the specific ethical risks defined in the AI governance framework are not included.

2) *Simple and clear*: to be practicable, classifications must be easy to understand and apply in practice. This implies that AI practitioners should be able to determine, with little effort, how to classify a specific use case. Ideally, the classification should be based on conditions that are discrete, i.e., which are either met or not. Finally, usefulness also implies that people without expert knowledge should be able to apply the classification.[8]

3) *Stable over time*: to allow for predictable governance, classifications must be resilient. Since AI research is a rapidly progressing field, AI systems' technical features and potential applications are subject to constant change. Hence, good classifications should not be based on elements that are likely to become obsolete too quickly.

Taken together, these criteria require that good models for classifying AI systems should help organisations specify to which systems or use cases their AI ethics principles apply, enable AI practitioners determine to which class of AI systems (if any) a particular use case belongs, and provide a stable basis for AI governance over time. These criteria also constitute the LoA on which we will evaluate the strengths and weaknesses of different models for classifying AI systems in Section 7.

Before proceeding, however, it is worth stressing that *how AI systems are classified* is an integral part of *the design of the AI governance framework* itself. For example, classifications that simply establish minimum thresholds for what constitutes AI systems demand flexible AI governance frameworks that can handle a wide range of use cases in proportionate and effective ways. In contrast, more fine-grained classifications afford layered AI governance frameworks that can specify both the risks and potential remedies associated with different use cases.

Building on this insight, it is possible to cluster pairs of *classifications of AI systems* and *AI governance frameworks* into three types. Using mental models, we have chosen to call these *the Switch*, *the Ladder*, and *the Matrix*. These are (of course) ideal types. In practice, many organisations use a combination of these approaches to demarcate the material scope of their AI governance frameworks. Nevertheless, as we shall see, the Switch, the Ladder and the Matrix are based on different logics. Hence, there is merit in describing, exemplifying, and evaluating them separately.

---

[8] This is because implementing any AI governance framework in practice requires the active participation of a wide range of staff across an organisation, including people who lack technical training and skills.

## 4   The Switch

Most AI governance frameworks include *working definitions* of AI systems. The aim is to capture the most relevant features of the systems under investigation in a single sentence or paragraph, providing policymakers and AI practitioners with a simple rule of thumb for when to apply the AI governance framework. Some of these working definitions are, as we shall see, too abstract to help establish a material scope. However, others are also useful for the practical purpose of implementing AI governance frameworks without complicating matters. Consider the approach taken by the IEEE. In 2020, the IEEE published its *Recommended Practice for Assessing the Impact of Autonomous and Intelligent Systems on Human Well-Being*. For the purpose of this framework,[9] the IEEE (2020) defined AI systems as follows:

> [An AI system is][10] … a semi-autonomous or autonomous computer-controlled system programmed to carry out some task with or without limited human intervention capable of decision making by independent inference and successfully adapting to its context.

This working definition highlights two central features of AI systems: the level of *autonomy* and the ability to *adapt*. Of course, both are a matter of degree. In some cases, AI systems act with complete autonomy, whereas in other cases, AI systems *only* provide recommendations to a human operator who has the final say (Cummings, 2004). Although it is a simplification, the IEEE's working definition is based on features that are directly linked to the specific ethical concerns posed by AI systems. It also enables AI practitioners to determine what is *not* within the AI governance framework's scope. For example, this definition does not cover non-adaptive expert systems that structure information for the convenience of human decision-makers.

The logic behind the IEEE's working definition of AI systems can be abstracted into what we call the Switch. The Switch is a model for binary classifications: something either *is* or *is not* an AI system. To establish such a threshold, the Switch consists of one or more *essential requirements*. These requirements can concern technical features (i.e., referring to what the system *is*) and functional aspects (i.e., referring to what the system *does*). Under ideal circumstances, simple yes/no questions are enough to determine whether a specific system satisfies the relevant requirement(s). Essential requirements are individually necessary and jointly sufficient. For our purposes, this means that an AI governance framework should apply to any use case that meets the requirements constituting the Switch. Figure 1 illustrates the logic behind the Switch approach.

---

[9] The IEEE different definitions of AI systems in different contexts. Here, we are talking explicitly about the definition included in IEEE 7010-2020, which was the first industry standard published in the field of AI ethics.
[10] Here, the IEEE does not use the term AI systems but rather Autonomous and Intelligent Systems (A/IS).

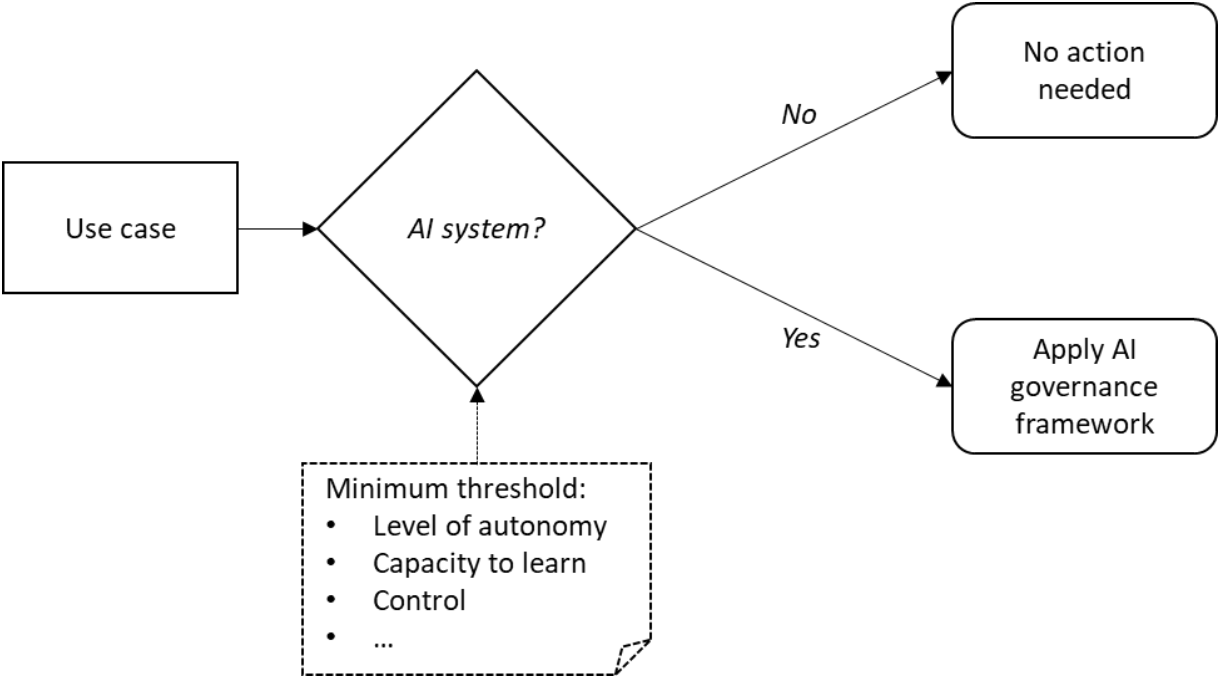

*Figure 1. The Switch – a binary approach to classifying AI systems according to some minimum threshold.*

The model presented above is a significant simplification. It does not describe exhaustively the process of determining the material scope for a specific AI governance framework. Nor does the model necessarily reflect the IEEE's intention behind their working definition of AI systems. Nevertheless, the model helps sketch the logic behind high-level, binary approaches to classifying AI systems for the purpose of AI governance. The IEEE's working definition of AI systems is a good example because it allows for meaningful evaluation of specific use cases. It is worth noticing that not all high-level working definitions of AI systems can serve the function of the Switch as described above. For example, in their report *Automating Society*, (AlgorithmWatch, 2019) takes a more holistic approach:

> [An AI system][11] … is a socio-technological framework that encompasses a decision-making model, an algorithm that translates this model into computable code, the data this code uses as an in-put—either to 'learn' from it or to analyse it by applying the model—and the entire political and economic environment surrounding its use.

The working definition of an AI system provided by AlgorithmWatch has many advantages. For example, a system-based approach requires that one examines the entire ecosystem in all of its complexities (Lauer, 2020). When technical subsystems are targeted separately, essential dynamics of the system as a whole may be lost or misunderstood (Di Maio, 2014). However, our aim here is not to compare the relative merits of different definitions of AI systems. Instead, we wish to exemplify how some high-level definitions of AI systems (the IEEE's included) can help organisations demarcate the material scope of the AI governance frameworks they seek to implement. In contrast, working definitions of AI systems that take '*the entire political and economic environment*' into account do not.

---

[11] AlgorithmWatch uses the term Automated Decision-Making (ADM).

In conclusion, the Switch is an intuitive model to classify AI systems for the purpose of AI governance. Its strength lies in the fact that it is easy for AI practitioners to remember and for organisations to communicate internally and externally. However, it is also a coarse approach that is likely to result in material scopes that are either under- or overinclusive. Hence, classifications of AI systems based on the Switch are feasible only in combination with flexible AI governance frameworks that, following an initial assessment, escalate only those use cases that demand further scrutiny.

## 5  The Ladder

A central function of AI governance frameworks is to put mechanisms in place that ensure accountability for AI systems and their outcomes, both before and after their implementation (European Commission, 2019). Because AI systems may exacerbate existing risks and introduce new ones, AI governance is closely linked to risk management, i.e., processes that allow different risks to be identified, understood, and managed (Leslie, 2019). According to ISO 31000 risk management guidelines (ISO 31000 - Risk Management - Guidelines, 2018), *risk* is the effect of uncertainty on objectives. So understood, risks can be ethical, legal, or technical. Faced with constant pressures to reduce uncertainty (Luhmann, 2018), most large organisations already have risk management frameworks in place (Currie, 2019). Hence, the use of AI systems does not necessarily require a complete overhaul of existing governance structures but rather an awareness of how AI systems may increase, or complicate the detection of, risks as they manifest themselves in unfamiliar ways (Lee et al., 2020). In short, successfully implementing AI governance entails the adoption of adequate measures to mitigate the ethical risks posed by a specific system in a manner proportionate to the magnitude of those risks (AI HLEG, 2019).

Building on this rationale, a growing number of proposals have advocated a risk-based approach to AI governance and hence to the classification of AI systems (Krafft et al., 2020).[12] Most notable amongst these proposals is the draft European regulation on AI, which takes an explicitly risk-based approach to AI governance (European Commission, 2020, 2021b). However, to exemplify the logic behind the approach, we will focus our analysis on the recommendation of the German Data Ethics Commission (DEK). The reason for this is that the DEK, as we shall see, outlined the risk-based approach to classifying AI systems in an outright and pedagogical manner.

In 2018, the DEK called for a risk-based approach to AI governance that would range from *no regulation* for the most innocuous AI systems to *a complete ban* for the most dangerous ones (DEK,

---

[12] In the US, most well-known example is the NIST AI risk management framework (NIST, 2022).

2018). In between these two extremes, the DEK defined three intermediary risk levels for which the use of AI systems is generally allowed but subjected to increasingly stringent governance requirements. Table 1 (below) summarises the five-level classification of AI systems for the purpose of AI governance as proposed by the DEK.[13]

*Table 1. The DEK's five-level classification of AI systems, based on their potential for harm.*

| Level | Potential for harm | Implications for AI governance |
|---|---|---|
| 1 | Applications with *zero* or *negligible* potential for harm | No special measures |
| 2 | Applications with *some* potential for harm | Measures such as formal and substantive requirements (e.g., transparency obligations) or monitoring procedures |
| 3 | Applications with *regular* or *significant* potential for harm | Additional measures such as ex-ante approval procedures |
| 4 | Applications with *serious* potential for harm | Additional measures such as a live interface for "always-on" oversight by supervisory institutions |
| 5 | Applications with an *untenable* potential for harm | Complete or partial ban of an algorithmic system |

The *potential for harm* is the key variable underpinning this classification of AI systems. The DEK suggested that the potential for harm can be determined by looking at the *likelihood* that harm will occur and the *severity* of that harm (DEK, 2018). Admittedly, determining the potential for harm is itself a non-trivial undertaking. We will revisit this question towards the end of this section. Here, it is sufficient to acknowledge that international standards have long been used to codify risk assessments successfully for many socio-technical issues, from cybersecurity to data privacy (ISO, 2019).

The risk-based approach taken by the DEK (and more generally by the EU) can be abstracted into a model for classifying AI systems for the purpose of AI governance. Here, we have chosen to call this model the Ladder. The basic idea behind the Ladder is to classify use cases into different risk levels. Depending on the ethical risks posed by a specific use case, different levels of AI governance apply. Of course, both the number of steps in the Ladder and the methods used to determine the risks posed by a specific system can vary between different contexts. Nevertheless, the basic principle remains simple: the greater the potential for harm, the more far-reaching the prescribed interventions. Figure 2 illustrates the generalisable logic behind the Ladder approach.

---

[13] The DEK uses the term 'algorithmic systems'.

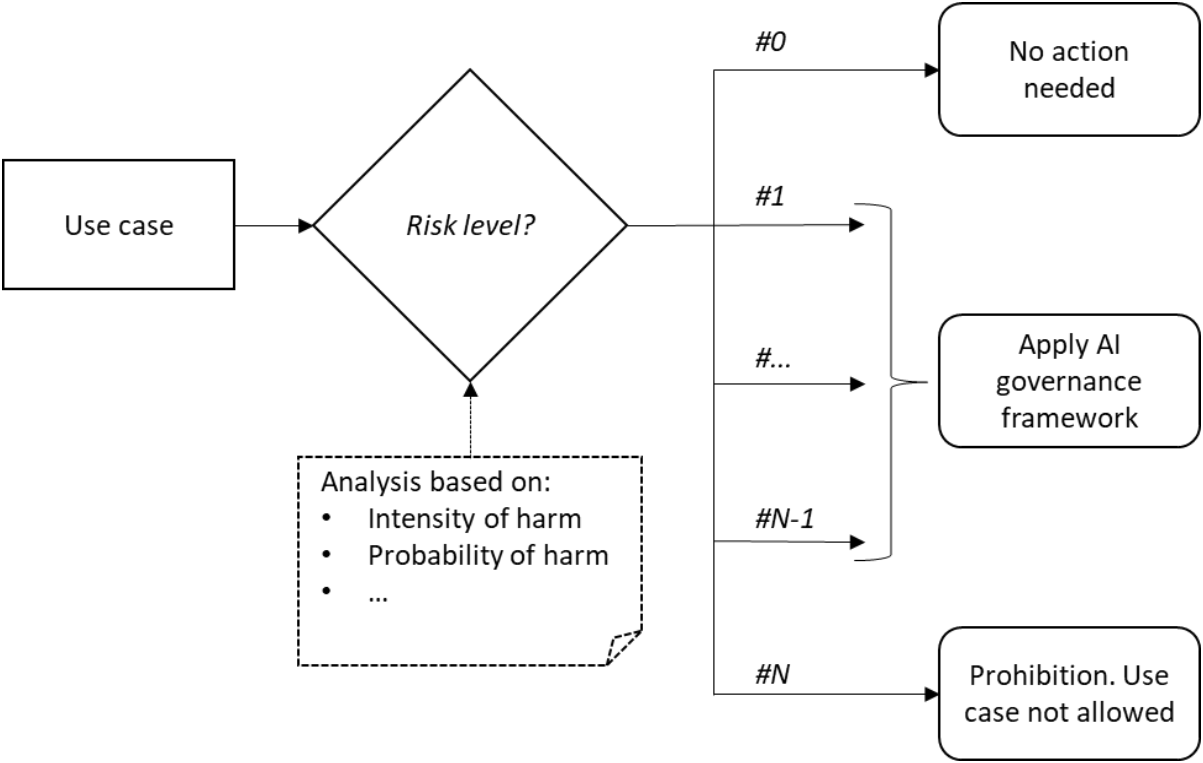

*Figure 2. The Ladder—a risk-based approach to classifying AI systems.*

Apart from the DEK and the European Commission, many other organisations have proposed similar risk-based approaches to AI governance. For example, the AI Ethics Impact Group (AIEIG), led by the VDE/Bertelsmann Stiftung, has developed a framework to help organisations operationalise AI ethics. The main idea behind their framework is to create a (standardised) context-independent labelling of AI systems inspired by the energy efficiency label (AIEIG, 2020). For their proposed AI Ethics Label, AIEIG assumes a set of ethical values, including accountability, transparency, and reliability. Each value is then broken down into criteria that define when values are honoured or violated. Significantly, these criteria rest on observables that can be monitored and quantified. This level of detail allows the AIEIG to be more specific than the DEK about the level of governance required for AI systems at each risk level. For example, the AIEIG distinguishes between the need for outcome accountability, objective-based accountability, and process accountability for AI systems at different intermediary risk levels.

But how are individual use cases classified into different risk levels? As mentioned above, the most common way to conceptualise risk is to combine the severity of harm and the likelihood of it occurring (Krafft et al., 2020). These two elements have previously been used to construct risk-ladders that allow for matching regulatory provisions to different risks, such as financial (MacNeil & O'Brien, 2010) or environmental risks (Black & Baldwin, 2012). The severity of harm typically depends on the task performed by the AI system and on what the possible outcomes are. For example, AI systems used for consumer recommendations may have a lower effect on human welfare than those used for job recruitment or medical interventions. In contrast, the likelihood of harm is often thought of in numerical terms as a probability. However, there is often uncertainty when evaluating real use cases, and exact probabilities are seldom known (Krafft et al., 2020). Hence, it is important to remember that the purpose of the Ladder, as outlined above, is not to identify hard thresholds between different risk levels but rather to provide practical guidance on how to classify AI systems in ways that are fit for purpose, simple and clear, and stable over time.

A further tension relates to the conflicting incentives actors within organisations may have when self-determining the risks associated with specific systems or use cases (The Government Office for Science, 2014). For example, executives facing pressure to cut costs and streamline the organisation may want to avoid that red flag are being raised too often. Similarly, individual managers or developers may prefer to handle the risks involved in a project locally, rather than escalating issues that might either kill the project or subject it to more administration and oversight.

To summarise, the Ladder is a purposeful and highly resilient model for classifying AI systems for the purpose of AI governance. By focusing on the risks posed by specific use cases, the Ladder presents a technology-agnostic framework that addresses the ethical risks posed by autonomous, self-learning systems. Its proven format also facilitates the integration of AI governance frameworks into existing organisational structures and processes. However, the Ladder requires AI practitioners to make an upfront risk assessment of individual use cases. This extra administrative burden – combined with the inherent uncertainty associated with the concept of risk – means that, compared with the simple Switch, the Ladder may be more costly and difficult to operationalise.

## 6   The Matrix

As discussed in Section 2, AI systems—although hard to define—typically share several characteristics. These characteristics include the ability to *perceive* the environment through input data, *process information* to interpret the data, *make decisions* based on the data, and, ultimately, *act* to achieve pre-defined goals (Samoili et al., 2020). Because AI systems are embedded in conditions that include and exceed them (Reddy et al., 2019), how they are used and their effects on society are not determined solely by the design of their internal decision-making models (Mökander & Axente, 2021). The extent to which an AI system's behaviour is perceived as ethical also depends on the input data, which could be incomplete or biased (Kim, 2017), and the context in which the system is applied (Lauer, 2020). Hence, the impact an AI system has on its environment is not always intuitive and may not be consistent over time. Rather, according to this perspective, every individual use case poses a unique set of ethical challenges.

To deal with this complexity, many organisations use multiple dimensions to classify AI systems. For example, the OECD (2022) has recently published a *Framework for the Classification of AI* *systems.*[14] The purpose this framework is to help organisations and policymakers assess and classify AI systems according to their potential ethical implications.

As a starting point, the OECD uses a socio-technical conceptualisation of AI systems. The OECD's framework thus requires organisations to analyse AI systems according to four key dimensions:

1)  Context: the socio-economic environment in which the AI system is deployed, notably the sector and the potential impact the system may have;

2)  Data and input: the data used by the AI model to build a representation of its environment;

3)  AI model: real-world processes in the system's internal environment, constituting the core of the AI system; and,

4)  Task and output: resulting actions taken by the system to influence its environment.

Each of these four key dimensions is broken down into subdimensions (see Table 2, below).

*Table 2. Dimensions and subdimensions of the OECD Framework for the Classification of AI Systems.*

| 1.  *Context* | 2.  *Data and input* | 3.  *AI model* | 4.  *Task and output* |
|---|---|---|---|
| Sector | Collection | AI model type | Task |
| Scale | Structure | AI model properties | Actions |
| Impact | Type | Acquisition of capabilities | Composite system |
| Critical | Quality | | Considerations |
| User | Collection | | |

Some of the subdimensions are binary. The *user* of the system, for example, is either an expert or a non-expert. Other subdimensions are categorical. One example here would be the *input data*, which may be structured, unstructured, or semi-structured. Yet other subdimensions, like explainability (Molnar, 2021; Rudin, 2019; Watson & Floridi, 2020), demand further clarification regarding their purpose and target audience and thus allow for free-form entries. As a whole, the framework attempts to capture all the most relevant aspects of AI systems to allow for effective governance.

The approach taken by the OECD represents a model to classify AI systems that we, in this article, have chosen to call the Matrix.[15] In essence, the Matrix refers to classifications of AI systems that take several different dimensions into account to identify the specific ethical risks associated with a particular use case. Of course, the number and scope of these dimensions can vary between different

---

[14] A first interim report was circulated in November 2020.
[15] Strictly speaking, a *matrix* is a special case of a *tensor*, i.e., a tensor of rank two. However, since higher order tensors are often represented as matrices, e.g., by column-binding data from later time points, we are not overly concerned with the distinction between these algebraic objects. In this article, we use the term *matrix* to refer to multidimensional spaces of arbitrary rank.

Matrix classification schemes. At least, these dimensions include the *technical characteristics* a system has, which *data* it can access, and the *context* in—and the purpose for which—the system is applied. For each combination in the Matrix, an AI governance framework can specify which preventive measures are needed to ensure that a particular AI system is designed and deployed in ethical ways. Figure 3 illustrates the logic behind the Matrix approach.

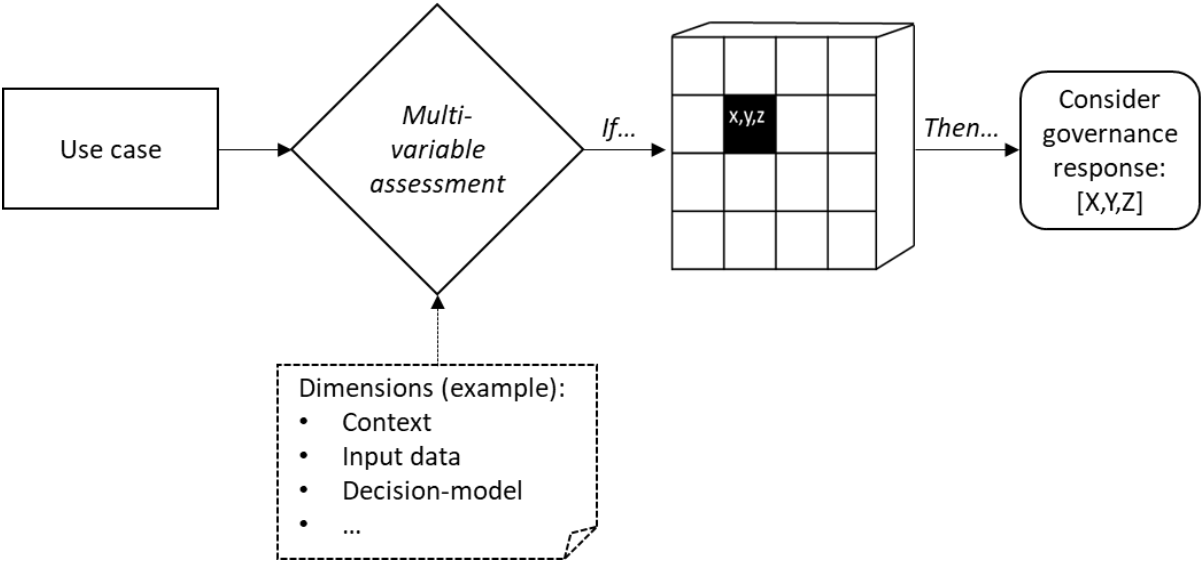

*Figure 3. The Matrix - a multi-dimensional classification of AI systems*

Due to this structure, the Matrix represents the most comprehensive, but also the most complicated, model for how to classify AI systems. To illustrate how quickly the model grows in complexity, let us consider how the context of an AI system can be assessed. The impact of a specific AI system may depend on several interrelated factors, including the breadth of deployment of the system, its maturity, the stakeholders with which it interacts, and the purpose for which it is applied. In theory, a broader deployment is coupled with greater risks than a narrower one, and a more mature AI system poses fewer risks than a less mature one. Further, the use of AI systems poses different ethical challenges in different sectors. Healthcare, for instance, is often considered a critical sector since even minor disruptions may have severe consequences for the health and safety of human beings. However, it is important to note that not all AI systems in a critical sector are critical. For example, a hospital's administrative time tracking systems need not necessarily be considered a critical system. This shows how evaluations in one dimension inevitably have implications for other dimensions in the Matrix.

In terms of input, good data governance demands that data sets are complete, accurate, and appropriate, whilst not being collected or processed in ways that infringe on individual privacy or intellectual property rights (ICO, 2018). However, concerning data management and information processing, the Matrix does need to not offer any new tools per se. Rather, it incorporates tools like model cards (Mitchell et al., 2019) and datasheets (Gebru et al., 2018; Holland et al., 2018) into a structured process for responding to the specific challenges associated with a given AI system. The key to this approach is the link between a particular use case's characteristics and the proposed remedies. For example, evaluating whether decisions made by AI systems are unfair or discriminatory typically use one, or a combination, of three distinct methods: *pre-processing*, *in-processing*, and *post-processing* (Clavell et al., 2020). Understanding about the way data is collected and structured informs pre-processing methods. In-processing, by contrast, depends on the AI model itself. Finally, post-processing techniques keep the AI model as it is but adjust the outcomes so that they correspond better to some predefined normative benchmark.

To summarise, the Matrix is the most comprehensive approach to classifying AI systems for the purpose of AI governance. The fact that it attempts to take all relevant information into account—from the type of decision-making model, which input data the model can access, and the context in which the system is applied—means that the Matrix can help organisations that design and deploy AI systems understand which precautions are necessary for a particular use case. At the same time, classifications based on the Matrix require substantial upfront investments, especially in terms of AI practitioners' time. Moreover, the different dimensions according to which AI systems are evaluated sometimes overlap and influence each other non-linearly. Hence, the Matrix is not a simple model to implement. Finally, its all-encompassing approach also makes the Matrix model less resilient to changes over time. In short, classifications based on the Matrix hold great promise to inform policy decisions. However, it may not necessarily help organisations define the material scope of their AI governance frameworks.

## 7   Discussion

Let us now evaluate the three models introduced in this article according to the criteria for good classifications of AI systems stipulated in Section 3. A significant advantage of the Switch is that it is simple and clear, i.e., easy to grasp and use even for non-experts. This is because a few essential requirements are easy to bear in mind yet still provide a meaningful basis for identifying the systems to which a specific AI governance framework applies. At the same time, the Switch may not always be fit for purpose, since all binary approaches to classifying AI systems—which attempt to draw a line across a continuous spectrum—struggle to strike the right balance between over- and under-inclusiveness. At first glance, it may appear better to err on the side of caution, i.e., to employ an overinclusive Switch. Doing so would increase the probability of identifying and managing high-risk use cases. At the same time, however, care must also be taken not to cause unjustifiable administrative burdens.[16] Finally, the extent to which a Switch is stable over time depends on which elements are included in the essential requirements. While requirements that refer to a particular design structure (e.g., neural networks) or a specific use case (e.g., facial recognition) are likely to change over time, essential requirements that refer to capabilities would be more permanent (Schuett, 2021).

Compared with the Switch, the Ladder has several conceptual advantages. First, it builds on well-established mechanisms like risk assessments and risk labels. This procedural continuity helps organisations integrate AI governance frameworks into their existing governance structures and quality management processes (Raji et al., 2020). Second, the Ladder is fit for purpose insofar as it demands that organisations assess the specific ethical challenges associated with the technical systems they design and the use cases for which these systems are deployed. This is important given that—even where they are technically similar—the consequences of the decisions informed or determined by AI systems may differ considerably depending on the concrete setting in which they are applied (Krafft et al., 2020). Finally, the fact that it is technology-agnostic makes the Ladder a highly stable model for classifying AI systems. As a result, AI governance frameworks that classify AI systems according to the Ladder are well equipped to handle both rapid technological innovation and social change. However, these advantages come at a cost. It can be very challenging to assess all the ethical risks posed by a specific use case in practice. Moreover, it remains challenging to quantify the externalities that occur due to indirect causal chains over time (Dafoe, 2017).

The Matrix is fit for purpose insofar as it informs the policy considerations associated with different types of AI systems. Classifications based on the Matrix help also help organisations identify which precautionary measures are appropriate when designing or implementing a specific AI system. However, the Matrix is not a simple model for classifying AI systems. In other words, it does not help AI practitioners to determine with little effort whether a particular use case should be subjected to the AI governance framework. Instead, the Matrix front-loads both the administrative burden—and the process of ethical deliberation—to the initial stages of software development processes. Finally, classifications based on the Matrix may not be stable over time. Many subdimensions consist of variables whose categories or ranges may change due to future technological advances. AI research is a quickly moving landscape and speculating about what technical breakthroughs may occur is beyond the scope of this article. The point is that any model that attempts to exhaust all possible combinations of available technologies, on the one hand, and potential areas of application, on the other, will struggle to stay relevant and useful.

---

[16] As a case in point, the Centre for Data Innovation recently estimated that 'the AIA will cost the European economy €31 bn over the next five years and reduce AI investments by almost 20 percent' (Mueller, 2021). How precise such numbers are is hard to say. Still, they serve as a healthy reminder that not only underinclusive but also overinclusive definitions of the material scope of AI governance are associated with real costs and risks.

Using a simple traffic-light assessment, Figure 4 below summarises the relative strengths and weaknesses of the Switch, the Ladder, and the Matrix as models to classify AI systems.

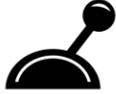

| | 'The Switch' | 'The Ladder' | 'The Matrix' |
|---|---|---|---|
| *Fit for purpose* | 🟡 | 🟡 | 🟢 |
| *Simple and clear* | 🟢 | 🟡 | 🔴 |
| *Stable over time* | 🟡 | 🟢 | 🟡 |

*Figure 4. Relative strengths and weaknesses of different models to classify AI systems*

Taking a step back, it should be re-emphasised that different ways of classifying AI systems can be combined and used by the same organisation at various stages of the governance process. An example of this possibility is found in the proposed European regulation on AI. As a first step, the AIA uses a Switch to demarcate its material scope. As mentioned in Section 2, the list of computational techniques coverer by the AIA is broad and include everything from logic-based to statistical approached. In a second step, however, the AIA relies on a risk-based approach, i.e., on the Ladder, to identify those AI systems that need to be subjected to additional oversight and transparency obligations.[17]

In practice, we expect that overlaps such as the one described above will be common for at least two reasons. First, because the Switch, the Ladder, and the Matrix have different (and complementary) strengths and weaknesses, organisations can tailor their material scope by combining the three approaches in various ways. Second, the three models compose a hierarchy of complexity in which each constitutes a special case of the previous: the Switch can be viewed as a bivalent Ladder, and the Ladder as a rank-one Matrix. Despite these overlaps, treating the three models for classifying AI systems as conceptually distinct is useful insofar as it enables a comparison of their respective affordances and constraints. In this article, we have therefore chosen to identify each model with the minimum level of complexity required to describe its operations – well aware of the fact that edge cases could strain the typology.

---

[17] This is not the place to discuss the merits and shortcomings of the material scope used in the European AIA. Readers interested in such a discussion are referred to Bryson (2022).

## 8 Conclusions

AI governance frameworks need to define their material scope if they are to make a difference in how organisations design and deploy AI systems. However, throughout this article, we have argued for a pragmatic approach: It is less important to define what AI is in abstract terms and more important to establish processes for identifying those systems or processes that require additional layers of governance. We have also shown that existing attempts to classify AI systems for the purpose of AI governance tend to follow one (or a combination) of the following three models: the Switch, the Ladder, or the Matrix.

All of these models come with their own set of strengths and weaknesses. The Switch is simple and clear—and can thus be easily communicated and applied. However, classifications of AI systems based on the Switch are often under- or overinclusive. In contrast, the Ladder provides a technology-neutral model for classifying AI systems that is fit for purpose and stable over time. However, although the risk-based approach, on which the Ladder is built, facilitates the integration of AI governance framework into existing governance structures, it also adds complexity that makes it more difficult to implement than the Switch. Finally, the Matrix offers the most comprehensive model for how to classify AI systems. As such, it is well-suited to inform policy decisions. However, classifications based on the Matrix add significant administrative burdens on organisations and AI practitioners and are also less stable over time.

In short, there is a three-way trade-off between how fit for purpose a model for classifying AI system is, how simple it is to apply, and how stable it is over time. When to apply classifications based on the Switch, the Ladder, or the Matrix thus remains a question to be evaluated locally, case by case. Moreover, it is important to remember that how best to classify AI systems depends on the design of the AI governance framework as a whole. Flexible and progressive AI governance frameworks must accompany simple classifications of AI systems. Highly detailed classifications of AI systems, on the other hand, allow for AI governance frameworks that proactively provide suggestions for how to manage the specific ethical challenges associated with a particular use case.

In this article, we have deliberately avoided any discussion about legal classifications of AI systems. However, it is worth stressing that the expectations on legal classifications of AI systems differ significantly from those intended for other purposes. Organisations that wish to improve their internal processes can afford some ambiguity concerning the material scope of their AI governance frameworks – as long as they operate within the space of legal permissibility and operational viability. In contrast, courts need to be able to determine precisely which systems or processes are affected by a specific law. We therefore anticipate the debate about how to classify AI system to intensify as the focus of the discourse concerning AI ethics increasingly shifts from 'soft' ethics principles to the emergence of 'hard' regulations (Floridi, 2018).

The analysis in this article has shown that it may neither be feasible nor desirable to define the material scope of AI governance frameworks at a cross-sectorial, technology neutral LoA. Rather than attempting to define AI, regulators should consider the role information processing play (and ought to play) in shaping social systems. This is partly because both human decision-makers and AI systems come with their own sets of strengths and weaknesses (Baum, 2017), and partly because ethical tensions do not emerge from the use of specific technologies alone but can also be intrinsic to the decision-making task (Danks & London, 2017). As a result, overly simplistic measures – such as introducing a blanket ban on specific technologies – are often unhelpful. A more fruitful approach would be to understand how different information processing systems shape their environments and subject the decision-making process itself to appropriate and proportionate quality assurance and transparency obligations.

To conclude, we hope that this article may serve as a map for those who seek to design or implement AI governance frameworks in practice. Our hope rests on the old idea that new insights can be gained through a multiplicity of perspectives. Models help us organise information to grasp complex phenomena, and the many-model approach helps illuminate the blind spots inherent in each model (Page, 2018). By drawing on the models outlined in this article, we hope that technology providers and regulators will be better equipped to find the classifications of AI systems that work best for them.

Finally, some may remain sceptical about whether it is necessary to classify AI systems at all. They may rightly point to the fact that AI systems are just a variation of other systems, with which they share many characteristics and ethical risks. However, human organisation is made possible by articulated conceptual representations of the world taken at a relatively *high level of abstraction* rather than the way the world *really is*. To quote John (Dewey, 1957) once more, "A classification is not a bare transcript or duplicate of some finished and done-for arrangement pre-existing in nature. It is rather a repertory of weapons for attack upon the future and the unknown."

This is a pre-print. A revised version has been accepted for publication in Minds and Machines.

## References

AI HLEG. (2019). *European Commission's Ethics Guidelines for Trustworthy Artificial Intelligence* (Issue May). https://ec.europa.eu/futurium/en/ai-alliance-consultation/guidelines/1

AI HLEG. (2020). *Assessment List for Trustworthy AI (ALTAI)*. https://ec.europa.eu/digital-single-market/en/news/assessment-list-trustworthy-artificial-intelligence-altai-self-assessment

AIEIG. (2020). From Principles to Practice - An interdisciplinary framework to operationalise AI ethics. *AI Ethics Impact Group, VDE Association for Electrical Electronic & Information Technologies e.V., Bertelsmann Stiftung*, 1–56. https://doi.org/10.11586/2020013

Aiken, C. (2021). *Classifying AI Systems CSET Data Brief*. https://cset.georgetown.edu/publication/classifying-ai-systems/

AlgorithmWatch. (2019). Automating society: Taking stock of automated decision-making in the EU. *Bertelsmann Stiftung*, 73–83. https://algorithmwatch.org/wp-content/uploads/2019/01/Automating_Society_Report_2019.pdf

Ananny, M., & Crawford, K. (2018). Seeing without knowing: Limitations of the transparency ideal and its application to algorithmic accountability. *New Media and Society*, *20*(3), 973–989. https://doi.org/10.1177/1461444816676645

AstraZeneca. (2020). *AstraZeneca data and AI ethics*. Position Statement. https://www.astrazeneca.com/sustainability/ethics-and-transparency/data-and-ai-ethics.html

Baldwin, R., & Cave, M. (1999). *Understanding regulation : theory, strategy, and practice*. Oxford University Press.

Baum, S. D. (2017). Social choice ethics in artificial intelligence. *AI and Society*, 1–12. https://doi.org/10.1007/s00146-017-0760-1

Black, J., & Baldwin, R. (2012). When risk-based regulation aims low: Approaches and challenges. *Regulation and Governance*, *6*(1), 2–22. https://doi.org/10.1111/j.1748-5991.2011.01124.x

BMW Group. (2020). *Seven principles for AI: BMW Group sets out code of ethics for the use of artificial intelligence*. Press Release. https://www.press.bmwgroup.com/global/article/detail/T0318411EN/seven-principles-for-ai:-bmw-group-sets-out-code-of-ethics-for-the-use-of-artificial-intelligence?language=en

Brundage, M., Avin, S., Wang, J., Belfield, H., Krueger, G., Hadfield, G., Khlaaf, H., Yang, J., Toner, H., Fong, R., Maharaj, T., Koh, P. W., Hooker, S., Leung, J., Trask, A., Bluemke, E., Lebensbold, J., O'Keefe, C., Koren, M., … Anderljung, M. (2020). Toward Trustworthy AI Development: Mechanisms for Supporting Verifiable Claims. *ArXiv*, *2004.07213[cs.CY])*. http://arxiv.org/abs/2004.07213

Bryson, J. (2022). *Europe Is in Danger of Using the Wrong Definition of AI | WIRED*. https://www.wired.com/story/artificial-intelligence-regulation-european-union/

This is a pre-print. A revised version has been accepted for publication in Minds and Machines.

Bryson, J., & Winfield, B. (2017). Standardizing Ethical Design for Artificial Intelligence and Autonomous Systems. *Computer*, *50*(5), 116–119. http://ieeexplore.ieee.org/stamp/stamp.jsp?tp=&arnumber=7924235

Burrell, J. (2016). How the machine 'thinks': Understanding opacity in machine learning algorithms. *Big Data & Society*, *3*(1). https://doi.org/10.1177/2053951715622512

Carnap, R. (1950). *Logical foundations of probability* [Book]. University of Chicago Press.

Cath, C., Cowls, J., Taddeo, M., & Floridi, L. (2018). Governing artificial intelligence: Ethical, legal and technical opportunities and challenges. *Philosophical Transactions of the Royal Society A: Mathematical, Physical and Engineering Sciences*, *376*(2133). https://doi.org/10.1098/rsta.2018.0080

Chen, M., & Golan, A. (2016). What May Visualization Processes Optimize? *IEEE Transactions on Visualization and Computer Graphics*, *22*(12), 2619–2632. https://doi.org/10.1109/TVCG.2015.2513410

Chopra, A. K., & Singh, M. P. (2018). Sociotechnical Systems and Ethics in the Large. *AIES 2018 - Proceedings of the 2018 AAAI/ACM Conference on AI, Ethics, and Society*, 48–53. https://doi.org/10.1145/3278721.3278740

Clavell, G. G., Zamorano, M. M. n., Castillo, C., Smith, O., & Matic, A. (2020). Auditing algorithms: On lessons learned and the risks of data minimization. *AIES 2020 - Proceedings of the AAAI/ACM Conference on AI, Ethics, and Society*, 265–271. https://doi.org/10.1145/3375627.3375852

Cookson, C. (2018, June 9). Artificial intelligence faces public backlash, warns scientist. *Financial Times*. https://www.ft.com/content/0b301152-b0f8-11e8-99ca-68cf89602132

Corea, F. (2019). *AI Knowledge Map: How to Classify AI Technologies*. 25–29. https://doi.org/10.1007/978-3-030-04468-8_4

Cruz Rivera, S., Liu, X., Chan, A. W., Denniston, A. K., Calvert, M. J., Darzi, A., Holmes, C., Yau, C., Moher, D., Ashrafian, H., Deeks, J. J., Ferrante di Ruffano, L., Faes, L., Keane, P. A., Vollmer, S. J., Lee, A. Y., Jonas, A., Esteva, A., Beam, A. L., … Rowley, S. (2020). Guidelines for clinical trial protocols for interventions involving artificial intelligence: the SPIRIT-AI extension. *Nature Medicine*, *26*(9), 1351–1363. https://doi.org/10.1038/s41591-020-1037-7

Cummings, M. L. (2004). Automation bias in intelligent time critical decision support systems. *Collection of Technical Papers - AIAA 1st Intelligent Systems Technical Conference*, *2*, 557–562.

Currie, N. (2019). Risk Based Approaches to Artificial Intelligence. In *Crowe Data Management*.

Cutler, A., Pribić, M., & Humphrey, L. (2018). Everyday Ethics for Artificial Intelligence. *Ibm*, 48. https://www.ibm.com/watson/assets/duo/pdf/everydayethics.pdf

Dafoe, A. (2017). AI Governance: A Research Agenda. In *American Journal of Psychiatry* (Issue July 2017). https://doi.org/10.1176/ajp.134.8.aj1348938

Danaher, J. (2012). *Is Technology Value-Neutral? New Technologies and Collective Action Problems*.

Danks, D., & London, A. J. (2017). Algorithmic bias in autonomous systems. *IJCAI International Joint Conference on Artificial Intelligence*, *0*(January), 4691–4697. https://doi.org/10.24963/ijcai.2017/654

de Laat, P. B. (2021). Companies Committed to Responsible AI: From Principles towards Implementation and Regulation? *Philosophy and Technology*, *34*(4), 1135–1193. https://doi.org/10.1007/s13347-021-00474-3

DEK. (2018). Opinion of the Data Ethics Commission. *Daten Ethik Kommssion*.

Deloitte. (2020). *Deloitte Introduces Trustworthy AI Framework to Guide Organizations in Ethical Application of Technology*. Press Release. https://www2.deloitte.com/us/en/pages/about-deloitte/articles/press-releases/deloitte-introduces-trustworthy-ai-framework.html

Dewey, J. (1957). *Reconstruction in philosophy* (Enl. ed. /). Beacon Press.

Di Maio, P. (2014). Towards a Metamodel to Support the Joint Optimization of Socio Technical Systems. *Systems*, *2*(3), 273–296. https://doi.org/10.3390/systems2030273

Erdelyi, O. J., & Goldsmith, J. (2018). Regulating Artificial Intelligence Proposal for a Global Solution. *AAAI/ACM Conference on Artificial Intelligence, Ethics and Society*. http://www.aies-conference.com/wp-content/papers/main/AIES_2018_paper_13.pdf

Eubanks, V. (2019). *Automating inequality : how high tools profile, police, and punish the poor* (First Picador edi...).

European Commission. (2019). *Communication: Building Trust in Human Centric Artificial Intelligence*.

European Commission. (2020). *White Paper On Artificial Intelligence-A European approach to excellence and trust*. 27.

European Commission. (2021a). *Proposal for REGULATION OF THE EUROPEAN PARLIAMENT AND OF THE COUNCIL* (COM(2021) 206 final).

European Commission. (2021b). *Proposal for REGULATION OF THE EUROPEAN PARLIAMENT AND OF THE COUNCIL* (COM(2021) 206 final).

Feigenbaum, E. A., & Feldman, J. (1963). *Computers and thought*. McGraw-Hill.

Ferretti, T. (2021). An Institutionalist Approach to AI Ethics: Justifying the Priority of Government Regulation over Self-Regulation. *Moral Philosophy and Politics*. https://doi.org/10.1515/mopp-2020-0056

Fjeld, jessica. (2020). Principled Artificial intelligence. *IEEE Instrumentation and Measurement Magazine*, *23*(3), 27–31. https://doi.org/10.1109/MIM.2020.9082795

Floridi, L. (2008). The method of levels of abstraction. *Minds and Machines*, *18*(3), 303–329. https://doi.org/10.1007/s11023-008-9113-7

Floridi, L. (2016). Faultless responsibility: On the nature and allocation of moral responsibility for distributed moral actions. *Philosophical Transactions of the Royal Society A: Mathematical, Physical and Engineering Sciences*, *374*(2083). https://doi.org/10.1098/rsta.2016.0112

Floridi, L. (2018). Soft Ethics and the Governance of the Digital. *Philosophy and Technology*, *31*(1). https://doi.org/10.1007/s13347-018-0303-9

Floridi, L. (2019). Translating Principles into Practices of Digital Ethics: Five Risks of Being Unethical. *Philosophy and Technology*, *32*(2), 185–193. https://doi.org/10.1007/s13347-019-00354-x

Floridi, L., & Cowls, J. (2019). A Unified Framework of Five Principles for AI in Society. *Harvard Data Science Review*, *1*, 1–13. https://doi.org/10.1162/99608f92.8cd550d1

Floridi, L., Cowls, J., Beltrametti, M., Chatila, R., Chazerand, P., Dignum, V., Luetge, C., Madelin, R., Pagallo, U., Rossi, F., Schafer, B., Valcke, P., & Vayena, E. (2018). AI4People—An Ethical Framework for a Good AI Society: Opportunities, Risks, Principles, and Recommendations. *Minds and Machines*, *28*(4), 689–707. https://doi.org/10.1007/s11023-018-9482-5

Frankish, K., & Ramsey, W. M. (2014). *The Cambridge handbook of artificial intelligence*.

Gasser, U., & Almeida, V. A. F. (2017). A Layered Model for AI Governance. *IEEE Internet Computing*, *21*(6), 58–62. https://doi.org/10.1109/MIC.2017.4180835

Gebru, T., Morgenstern, J., Vecchione, B., Vaughan, J. W., Wallach, H., Daumé, H., & Crawford, K. (2018). *Datasheets for Datasets*. http://arxiv.org/abs/1803.09010

Goldstein, B. (2018). *A Brief Taxonomy of AI*. *March*, 1–12.

Google. (2018). *Artificial intelligence at Google: Our principles*. Communication. https://ai.google/principles/

Grant, M. J., & Booth, A. (2009). A typology of reviews: An analysis of 14 review types and associated methodologies. *Health Information and Libraries Journal*, *26*(2), 91–108. https://doi.org/10.1111/j.1471-1842.2009.00848.x

Grote, T., & Berens, P. (2020). On the ethics of algorithmic decision-making in healthcare. *Journal of Medical Ethics*, *46*(3), 205–211. https://doi.org/10.1136/medethics-2019-105586

Hastie, T., Tibshirani, R., & Friedman, J. H. (2009). *The elements of statistical learning : data mining, inference, and prediction* (2nd ed.). Springer.

Hodges, C. (2015). Ethics in business practice and regulation. *Law and Corporate Behaviour : Integrating Theories of Regulation, Enforcement, Compliance and Ethics*, 1–21. https://doi.org/10.5040/9781474201124

Holland, S., Hosny, A., Newman, S., Joseph, J., & Chmielinski, K. (2018). *The Dataset Nutrition Label: A Framework To Drive Higher Data Quality Standards*. *May*. http://arxiv.org/abs/1805.03677

Ibáñez, J. C., & Olmeda, M. V. (2021). Operationalising AI ethics: how are companies bridging the gap between practice and principles? An exploratory study. *AI and Society*. https://doi.org/10.1007/s00146-021-01267-0

ICO. (2018). Guide to the General Data Protection Regulation (GDPR). *Guide to the General Data Protection Regulation*, n/a. https://doi.org/10.1111/j.1751-1097.1994.tb09662.x

IEEE. (2019). Ethically aligned design. *Intelligent Systems, Control and Automation: Science and Engineering*, *95*, 11–16. https://doi.org/10.1007/978-3-030-12524-0_2

IEEE. (2020). *IEEE Standards Dictionary Online*. http://dictionary.ieee.org

IEEE SA. (2020). *IEEE 7010-2020 - IEEE Recommended Practice for Assessing the Impact of Autonomous and Intelligent Systems on Human Well-Being*.

ISO 31000 - Risk Management - Guidelines, Pub. L. No. ISO 31000:2018 (2018).

ISO. (2019). *It's all about trust*. ISO News. https://www.iso.org/news/ref2452.html

Jobin, A., Ienca, M., & Vayena, E. (2019). *Artificial Intelligence: the global landscape of ethics guidelines*.

Johnson-Laird, P. N. (1983). *Mental models : towards a cognitive science of language, inference, and consciousness*. Cambridge University Press.

Kahneman, D. (2011). Thinking, fast and slow [Article]. *Fortune*, *172*(1), 20–20.

Kim, P. (2017). Auditing Algorithms for Discrimination. *University of Pennsylvania Law Review*, *166*, 189–203.

Kostopoulos, L. (2021). Decoupling Human Characteristics from Algorithmic Capabilities. *The IEEE Standard Association*.

Krafft, T. D., Zweig, K. A., & König, P. D. (2020). How to regulate algorithmic decision-making: A framework of regulatory requirements for different applications. *Regulation and Governance*, *October*. https://doi.org/10.1111/rego.12369

Kritikos, M. (2019). AI Briefing: Legal & ethical reflections. *Briefing. European Parliament*, *March*.

Lauer, D. (2020). You cannot have AI ethics without ethics. *AI and Ethics*, *0123456789*, 1–5. https://doi.org/10.1007/s43681-020-00013-4

Lee, M., Floridi, L., & Denev, A. (2020). Innovating with confidence: Embedding Governance and Fairness in a Financial Services Risk Management Framework. *Berkeley Technology Law Journal*.

This is a pre-print. A revised version has been accepted for publication in Minds and Machines.

Legg, S., & Hutter, M. (2007). *A Collection of Definitions of Intelligence*. 1–12. http://arxiv.org/abs/0706.3639

Leslie, D. (2019). *Understanding artificial intelligence ethics and safety: A guide for the responsible design and implementation of AI systems in the public sector.* 97. https://doi.org/10.5281/zenodo.3240529

Liu, X., Cruz Rivera, S., Moher, D., Calvert, M. J., & Denniston, A. K. (2020). Reporting guidelines for clinical trial reports for interventions involving artificial intelligence: the CONSORT-AI extension. *Nature Medicine*, *26*(9), 1364–1374. https://doi.org/10.1038/s41591-020-1034-x

Luhmann, N. (2018). *Organization and decision*.

Marcus, G. (2020). The next decade in AI: Four steps towards robust artificial intelligence. *ArXiv*, *February*.

McCarthy, J. (2007). What is artificial intelligence? *Stanford University*.

Microsoft. (2019). *Microsoft AI Principles*. Communication. https://www.microsoft.com/en-us/ai/our-approach-to-ai

Mitchell, M., Wu, S., Zaldivar, A., Barnes, P., Vasserman, L., Hutchinson, B., Spitzer, E., Raji, I. D., & Gebru, T. (2019). Model cards for model reporting. *FAT\* 2019 - Proceedings of the 2019 Conference on Fairness, Accountability, and Transparency*, *Figure 2*, 220–229. https://doi.org/10.1145/3287560.3287596

Mökander, J., & Axente, M. (2021). Ethics-Based Auditing of Automated Decision-Making Systems : Intervention Points and Policy Implications. *AI & SOCIETY*, *0123456789*. https://doi.org/10.1007/s00146-021-01286-x

Mökander, J., Axente, M., Casolari, F., & Floridi, L. (2021b). Conformity Assessments and Post-market Monitoring : A Guide to the Role of Auditing in the Proposed European AI Regulation. *Minds and Machines*, 1–27.

Mökander, J., & Floridi, L. (2022). Operationalising AI governance through ethics-based auditing: an industry case study. *AI and Ethics*. https://doi.org/10.1007/s43681-022-00171-7

Mökander, J., Morley, J., Taddeo, M., & Floridi, L. (2021a). Ethics-based auditing of automated decision-making systems: nature, scope, and limitations. *Science and Engineering Ethics*, 1–30. https://doi.org/10.1007/s11948-021-00319-4 ORIGINAL

Molnar, C. (2021). Interpretable Machine Learning. A Guide for Making Black Box Models Explainable. *Book*, 247.

Morley, J., Floridi, L., Kinsey, L., & Elhalal, A. (2020). From What to How: An Initial Review of Publicly Available AI Ethics Tools, Methods and Research to Translate Principles into Practices [Article]. *Science and Engineering Ethics*, *26*(4), 2141. https://doi.org/10.1007/s11948-019-00165-5

Morley, J., Kinsey, L., Elhalal, A., Garcia, F., Ziosi, M., Floridi, L., & Morley, J. (2021). Operationalising AI ethics : barriers , enablers and next steps. *AI & SOCIETY, Villarreal 2020*. https://doi.org/10.1007/s00146-021-01308-8

Mueller, B. (2021). *How Much Will the Artificial Intelligence Act Cost Europe? July*.

NIST. (2022, April 12). *AI Risk Management Framework | NIST*. https://www.nist.gov/itl/ai-risk-management-framework

OECD. (2019). Recommendation of the Council on Artificial Intelligence. *OECD/LEGAL/0449*.

OECD. (2020). OECD Framework for the Classification of AI Systems – Interim report on the activities of the ONE AI Working Group on the Classification of AI Systems. *DSTI/CDEP(2020)13, November*.

OECD. (2022). *OECD Framework for the Classification of AI systems*. https://doi.org/10.1787/cb6d9eca-en.

Office of U.S. Senator Ron Wyden. (2022). Algorithmic Accountability Act of 2022. *117th Congress 2D Session*. https://doi.org/10.1016/S0140-6736(02)37657-8

Oxborough, C., Cameron, E., Rao, A., Birchall, A., Townsend, A., & Westermann, C. (2018). *Explainable AI*. https://www.pwc.co.uk/audit-assurance/assets/explainable-ai.pdf

Page, S. E. (2018). *The Model Thinker: What You Need to Know to Make Data Work for You*. Basic Books, Inc.

PwC. (2019). *A practical guide to Responsible Artificial Intelligence ( AI )*. https://www.pwc.com/gx/en/issues/data-and-analytics/artificial-intelligence/what-is-responsible-ai/responsible-ai-practical-guide.pdf

Rahwan, I. (2018). Society-in-the-loop: programming the algorithmic social contract. *Ethics and Information Technology, 20*(1), 5–14. https://doi.org/10.1007/s10676-017-9430-8

Raji, I. D., Smart, A., White, R. N., Mitchell, M., Gebru, T., Hutchinson, B., Smith-Loud, J., Theron, D., & Barnes, P. (2020). Closing the AI accountability gap: Defining an end-to-end framework for internal algorithmic auditing. *FAT* 2020 - Proceedings of the 2020 Conference on Fairness, Accountability, and Transparency*, 33–44. https://doi.org/10.1145/3351095.3372873

Reddy, E., Cakici, B., & Ballestero, A. (2019). Beyond mystery: Putting algorithmic accountability in context. *Big Data and Society, 6*(1), 1–7. https://doi.org/10.1177/2053951719826856

Reisman, D., Schultz, J., Crawford, K., & Whittaker, M. (2018). Algorithmic impact assessments: A practical framework for public agency accountability. *AI Now Institute, April*, 22. https://ainowinstitute.org/aiareport2018.pdf

Rolnick, D., Donti, P. L., Kaack, L. H., Kochanski, K., Lacoste, A., Sankaran, K., Ross, A. S., Milojevic-Dupont, N., Jaques, N., Waldman-Brown, A., Luccioni, A., Maharaj, T., Sherwin, E. D., Karthik

Mukkavilli, S., Kording, K. P., Gomes, C., Ng, A. Y., Hassabis, D., Platt, J. C., … Bengio, Y. (2019). Tackling climate change with machine learning. *ArXiv*.

Rudin, C. (2019). Stop explaining black box machine learning models for high stakes decisions and use interpretable models instead. *Nature Machine Intelligence*, *1*(5), 206–215. https://doi.org/10.1038/s42256-019-0048-x

Russell, S., Dewey, D., & Tegmark, M. (2015). *Artificial Intelligence*. 105–114.

Russell, S. J., & Norvig, P. (2015). *Artificial intelligence : a modern approach* (Third edit).

Saleiro, P., Kuester, B., Hinkson, L., London, J., Stevens, A., Anisfeld, A., Rodolfa, K. T., & Ghani, R. (2018). *Aequitas: A Bias and Fairness Audit Toolkit*. *2018*. http://arxiv.org/abs/1811.05577

Samoili, S., López Cobo, M., Gómez, E., De Prato, G., Martínez-Plumed, F., & Delipetrev, B. (2020). *AI Watch. Defining Artificial Intelligence. Towards an operational definition and taxonomy of artificial intelligence*. https://doi.org/10.2760/382730

Sandvig, C., Hamilton, K., Karahalios, K., & Langbort, C. (2014). Auditing Algorithms. *ICA 2014 Data and Discrimination Preconference*, 1–23. https://doi.org/10.1109/DEXA.2009.55

Scherer, M. (2016). Regulating artificial intelligence systems: Risks, challenges, competences, and strategies. *Harvard Journal of Law & Technology*, *29*(2), 98. https://doi.org/10.1007/s00521-010-0388-2

Schiff, D., Borenstein, J., Biddle, J., & Laas, K. (2021a). AI Ethics in the Public, Private, and NGO Sectors: A Review of a Global Document Collection. *IEEE Transactions on Technology and Society*, *2*(1), 31–42. https://doi.org/10.1109/tts.2021.3052127

Schiff, D., Rakova, B., Ayesh, A., Fanti, A., & Lennon, M. (2021b). Explaining the Principles to Practices Gap in AI. *IEEE Technology and Society Magazine*, *40*(2), 81–94. https://doi.org/10.1109/MTS.2021.3056286

Schneider, G. (2019). Mind and machine in drug design. *Nature Machine Intelligence*, *1*(3), 128–130. https://doi.org/10.1038/s42256-019-0030-7

Schuett, J. (2021). Defining the scope of AI regulations. *ArXiv*. http://arxiv.org/abs/1909.01095

Silver, D., Huang, A., Maddison, C. J., Guez, A., Sifre, L., van den Driessche, G., Schrittwieser, J., Antonoglou, I., Panneershelvam, V., Lanctot, M., Dieleman, S., Grewe, D., Nham, J., Kalchbrenner, N., Sutskever, I., Lillicrap, T., Leach, M., Kavukcuoglu, K., Graepel, T., & Hassabis, D. (2016). Mastering the game of Go with deep neural networks and tree search. *Nature*, *529*(7587), 484–489. https://doi.org/10.1038/nature16961

Smith, B. C. (2019). *The promise of artificial intelligence : reckoning and judgment*.

Sutton, R. S., & Barto, A. G. (2018). *Reinforcement learning : an introduction* (Second edi).

This is a pre-print. A revised version has been accepted for publication in Minds and Machines.

Taddeo, M., & Floridi, L. (2018). How AI can be a force for good. *Science*, *361*(6404), 751–752. https://doi.org/10.1126/science.aat5991

Tam, G. K. L., Kothari, V., & Chen, M. (2017). An Analysis of Machine- and Human-Analytics in Classification. *IEEE Transactions on Visualization and Computer Graphics*, *23*(1), 71–80. https://doi.org/10.1109/TVCG.2016.2598829

The Government Office for Science. (2014). *INNOVATION: MANAGING RISK, NOT AVOIDING IT - Evidence and Case Studies*.

Tsamados, A., Aggarwal, N., Cowls, J., Morley, J., Roberts, H., Taddeo, M., & Floridi, L. (2020). The Ethics of Algorithms: Key Problems and Solutions. *SSRN Electronic Journal*, *August*. https://doi.org/10.2139/ssrn.3662302

*US National Defence Authorization Act*, (2018) (testimony of USDOD). https://doi.org/H.R.2810

van de Poel, I. (2020). Embedding Values in Artificial Intelligence (AI) Systems. *Minds and Machines*, *30*(3), 385–409. https://doi.org/10.1007/s11023-020-09537-4

van Leeuwen, J. (2014). On Floridi's method of levels of abstraction. *Minds and Machines*, *24*(1), 5–17. https://doi.org/10.1007/s11023-013-9321-7

Vanschoren, J. (2018). *Meta-Learning: A Survey*. 1–29.

Vinuesa, R., Azizpour, H., Leite, I., Balaam, M., Dignum, V., Domisch, S., Felländer, A., Langhans, S., Tegmark, M., & Nerini, F. F. (2019). *The role of artificial intelligence in achieving the Sustainable Development Goals*. https://doi.org/10.1038/s41467-019-14108-y

Wang, P. (2019). On Defining Artificial Intelligence. *Journal of Artificial General Intelligence*, *10*(2), 1–37. https://doi.org/10.2478/jagi-2019-0002

Watson, D. S., & Floridi, L. (2020). The explanation game: a formal framework for interpretable machine learning. In *Synthese*. Springer Netherlands. https://doi.org/10.1007/s11229-020-02629-9

Weiss, K., Khoshgoftaar, T. M., & Wang, D. D. (2016). A survey of transfer learning. In *Journal of Big Data* (Vol. 3, Issue 1). Springer International Publishing. https://doi.org/10.1186/s40537-016-0043-6

Whittaker, M., Crawford, K., Dobbe, R., Fried, G., Kaziunas, E., Mathur, V., Myers West, S., Richardson, R., Schultz, J., & Schwartz, O. (2018). AI Now Report 2018. *AI Now*, *December*, 1–62.

Whittlestone, J., Alexandrova, A., Nyrup, R., & Cave, S. (2019). The role and limits of principles in AI ethics: Towards a focus on tensions. *AIES 2019 - Proceedings of the 2019 AAAI/ACM Conference on AI, Ethics, and Society*, 195–200. https://doi.org/10.1145/3306618.3314289